\documentstyle[aps,prl,epsfig]{revtex}

\newcommand{\be}{\begin{equation}}
\newcommand{\ee}{\end{equation}}
\newcommand{\bd}{\begin{displaymath}}
\newcommand{\ed}{\end{displaymath}}
\newcommand{\ba}{\begin{eqnarray}}
\newcommand{\ea}{\end{eqnarray}}
\newcommand{\bi}{\begin{itemize}}
\newcommand{\ei}{\end{itemize}}

\newcommand{\sbbox}[1]{\mbox{\scriptsize\bf $#1$}}

\begin{document}
\draft

\twocolumn[\hsize\textwidth\columnwidth\hsize\csname
@twocolumnfalse\endcsname

\preprint{PACS: 05.70.Jk, 11.10.Wx, 98.80.Cq,
DFTUZ/97/27\\
{}~hep-lat/yymmnn}

\title{A Monte Carlo
study of Inverse Symmetry Breaking. }
\author{G. Bimonte$^*$},
\address{\it  Dipartimento di Scienze Fisiche, Universit\`{a} di Napoli,
Mostra d'Oltremare, Pad.19, I-80125, Napoli, Italy}
\author{D. I\~{n}iguez, A. Taranc\'{o}n  and
C.L.Ullod
$^{\dagger}$
}
\address{\it  Departamento de F\'{\i}sica Te\'{o}rica, Facultad de Ciencias,
Universidad de Zaragoza, 50009 Zaragoza, Spain }
\date{February, 1998}
\maketitle

\begin{abstract}

We make a Monte Carlo study of the coupled two-scalar
$\lambda\phi^2_1\phi^2_2$ model in four dimensions at finite temperature.
We find no trace of Inverse Symmetry Breaking for values of the
renormalized parameters for which perturbation theory predicts this
phenomenon.

\end{abstract}
\vskip1pc]

Is it always true that more heat means more disorder? The intuitive answer
to this question would be yes and indeed this is what happens in the
majority of physical systems. Nevertheless, over 20 years ago, S. Weinberg
\cite{wei}, quoting an unpublished remark by S. Coleman, observed that
there may be exceptions to the general rule: in models with a sufficiently
rich scalar sector, some of the scalars may acquire a {\it negative} Debye
mass, with the result that the symmetric vacuum becomes necessarily
unstable at high temperatures. This remarkable phenomenon is called
Inverse-Symmetry-Breaking (ISB) or Symmetry-Non-Restoration (SNR) depending
on the symmetry of the ground state at zero temperature. The possibility of
ISB and SNR in realistic particle models and their cosmological
consequences have been explored in a number of papers \cite{ms}. Weinberg's
analysis of ISB and SNR  was based on a simple one-loop approximation and
some authors have questioned its reliability. Subsequent studies, using
different approximations with various amounts of non-perturbative content,
produced contradictory results. While some concluded that ISB and SNR
cannot occur \cite{fuj}, others found that they do occur \cite{bim}, even
though (in the majority of the cases) in a region of the parameter space
significantly reduced with respect to the one-loop result .

In this paper we present the first Monte Carlo study of ISB
in 4 dimensions in a two-scalar model, with
a global ${\bf Z}_2 {\times} {\bf Z}_2$ symmetry.
The job was mainly carried out on our RTNN computer, which holds
32 PentiumPro processors, for a total CPU time of approximately two
months of the whole machine.
The results were analogous to those found in
2+1 dimensions \cite{bitu} and seem to
show that ISB is absent for certain values of the renormalized
couplings in the region for which perturbation theory predicts it,
in accordance with the theorem proven in \cite{bim2}.

The model we simulated is described by the bare (euclidean) action:
$$
S=\int d^4 x
\Big\{ \sum_{i=1,2}\Big[
\frac{1}{2}(\partial_{\mu}\Phi_i^{(0)2}) + \frac{1}{2}m_i^{(0)2}
\Phi_i^{(0)2} + \frac{g_i^{(0)}}{4!} \Phi_i^{(0)4}\Big]
$$
\be
~~~~~+\frac{1}{4} g^{(0)}\Phi_1^{(0)2}\Phi_2^{(0)2} \Big\}.
\label{bact1}
\ee
If $g^{(0)}<0$ the potential is bounded from below for:
\ba
g_1^{(0)} \; g_2^{(0)} > 9\; g^{(0)2},~~~g_1^{(0)}> 0,~~~g_2^{(0)}> 0.
\label{cotas}
\ea
At high temperature, according to a simple one loop computation \cite{wei},
the thermal mass $M_i^2(T)$ is proportional to $T^2(g^R_i + g^R)$, where
$g_i^R$ and $g^R$ are renormalized parameters. ISB (or SNR)  occurs when
the coupling among the two fields $g^R$ is {\it negative} and such that,
say,
\be
g_2^R + g^R < 0.
\label{inst}
\ee
Then, $M_2^2(T)$ becomes negative at high T and this means that if one
starts from the disordered phase at $T=0$, the field $\Phi_2^{(0)}$ should
get ordered at sufficiently high temperature. This is the essence of ISB
and what we have tried to investigate by means of a Monte Carlo simulation.

As is customary on the lattice, we rewrite the action (\ref{bact1}) in
terms of dimensionless quantities as:
\ba
S_{L}&&=  \sum_{\sbbox{r} \in Z^4} \Big\{ \sum_{i=1,2} \Big[
       -\kappa_i  \sum_{\mu}\phi_{i,\sbbox{r}} \phi_{i,\sbbox{r}+
\hat{\sbbox{\mu}}}+
       \lambda_i(\phi^2_{i,\sbbox{r}}-1)^2 \nonumber\\
&&~~~~~~~~~~~~~~~~~~+{\phi}^2_{i,\sbbox{r}} \Big]+
\lambda\phi^2_{1,\sbbox{r}}\phi^2_{2,\sbbox{r}}\Big\},
\label{bbfin}
\ea
where
\ba
g_i^{(0)}=\frac{24 \lambda_i}{\kappa_i^2},~~ a^2
m_i^{(0)2}=2\frac{1-2\lambda_i -4 \kappa_i}{\kappa_i},~~ g^{(0)}=\frac{4
\lambda}{\kappa_1\kappa_2}.
\ea
In principle, one would then proceed as follows.  First, one would draw the
phase diagram of the system (\ref{bbfin}) at $T=0$. The physics described
by renormalized perturbation theory would be recovered in the scaling
region of this phase diagram, near the surface where both fields become
critical. In order to check the prediction of ISB made by perturbation
theory, one would then pick a point (the $T=0$ theory) in the disordered
phase, well inside the scaling region. That point should be chosen such
that the renormalized couplings were as small as possible (for perturbation
theory to be reliable) and verified the one-loop condition (\ref{inst}) for
ISB as strongly as possible. For any choice of the bare parameters, the
first condition is always satisfied if one  goes deep enough in the scaling
region, because the model is expected to be trivial. As for the second, it
should be achieved by a suitable choice of the bare parameters. Having
chosen the $T=0$ theory, one would then study the effects of temperature by
simulating the model on lattices having a finite extension $N_t$ in the
euclidean time direction. The physical temperature is related to $N_t$ as
$T=1/(a N_t)$. For ISB to occur it should happen that the point for the
$T=0$ theory selected above lied in the ordered region with
$\langle\phi_2\rangle\neq 0$ of all the phase diagrams with an $N_t$
smaller than some $\bar N_t$. On the contrary, the existence of ISB would
be ruled out if one found that the size of the disordered region
monotonically increases when the temperature is raised, namely when $N_t$
is reduced, starting from $N_t=\infty$, as it happens in normal systems,
see for example \cite{jan}.

In practice, it is impossible to realize this program, because drawing
five-dimensional phase diagrams with the high level of numerical precision
that turned out to be necessary would have required an enormous time. Thus,
we decided to fix once and for all the values of $\lambda_1$, $\lambda_2$
and $\lambda$ and then study the phase diagrams in the $\kappa_1,\kappa_2$
plane. We took $\lambda_1=0.3375$, $\lambda_2=0.01125$. The condition of
stability for the bare potential, eq.(\ref{cotas}), then gave a lower bound
for $\lambda$ of $-0.123$, and for our simulations we  selected the value
$\lambda=-0.112$ which is very close to the instability bound, thus giving
us the best chances of observing ISB.

\begin{figure}[h!]
\epsfig{figure= 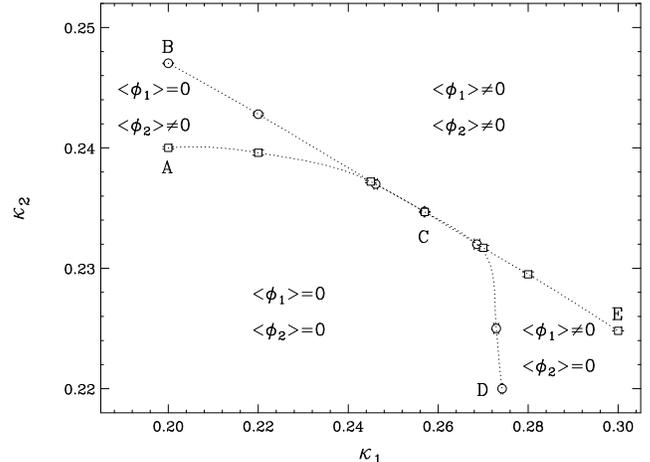,angle=90,width=240pt}
\caption{Phase diagram at $\lambda_1=0.3375$,
$\lambda_2=0.01125$, $\lambda=-0.112$ for a $4^4$ lattice.}
\label{PH_DIA}
\end{figure}

The corresponding $T=0$ phase diagram, obtained on a single $4^4$ lattice,
is shown in fig.\ref{PH_DIA}. We observe that the $(\kappa_1,\kappa_2)$
plane is divided in four regions. The lower left region is the disordered
phase, while the upper right one is the totally ordered phase. The wedges
between them represent partially ordered phases: in the upper left wedge
the field $\phi_2$ is ordered, while $\phi_1$ is disordered, while in the
lower right wedge it occurs the contrary. The critical line $A$ is of
second order, and we have obtained gaussian exponents. We have not studied
accurately the thermodynamical limit of the $B$, $D$ and $E$ lines, but we
saw a clear evidence of second order behavior on $D$ and strong first order
transitions across $B$ and $E$. The critical lines cross at a point C,
whose neighbourhood contains the scaling region because C evolves towards
the gaussian fixed point when $\kappa_i \to 1/4$ and $\lambda_i,\lambda \to
0$. Probing this region turned out to be practically impossible: while on
small lattices it was not even possible to distinguish the various
transition lines near $C$, on large ones the strong metastability due to
the first order transitions prevented us from getting clean results.

The next step was to check if in the scaling region of the disordered phase
there were points where the renormalized couplings satisfied the
perturbative condition for ISB. We thus selected a point
($\kappa_1=0.255,\kappa_2=0.234$) not far from the critical point C
and measured there the renormalized couplings and the correlation lengths
for both fields.
For these measurements we used $N_s^4$ lattices, with $N_s=6,8,10$,
making between 50 and 250 million iterations, with
autocorrelation times of some tens of iterations.

The renormalized couplings were measured using the following estimators:
\ba
g_i^R(N_s,\kappa_1,\kappa_2)= 2 U_i\frac{N_s^4}{\xi_i^4} \ \ \ i=1,2
\ea
\ba
{g}^R(N_s,\kappa_1,\kappa_2)= U_{12}\frac{N_s^4}{{\xi_1}^{2}{\xi_2}^{2}}
\ea
where $\xi_i(N_s,\kappa_1,\kappa_2)$ is the second moment
correlation length \cite{bitu}. $U_i(N_s,\kappa_1,\kappa_2)$ and
$U_{12}(N_s,\kappa_1,\kappa_2)$ are the Binder cumulants:
\be
U_i(N_s,\kappa_1,\kappa_2)=\frac{3}{2}-
\frac{\langle M_i^4 \rangle}{2\langle M_i^2\rangle^2}~,\label{bin}
\ee
\be
U_{12}(N_s,\kappa_1,\kappa_2)=1-
\frac{\langle M_1^2 M_2^2 \rangle}{\langle M_1^2\rangle\langle M_2^2\rangle}~,
\label{bin12}
\ee
where $M_i$ is the average magnetization
\be
M_i= \frac{1}{V}\sum_{\sbbox{r}}\phi_{i,\sbbox{r}}~.
\ee
\begin{table}[h!]
%\begin{center}
\begin{tabular}{|c|c|c|c|c|c|}
\hline
$N_s$ & $\xi_1$ & $\xi_2$ & $g_1^R$ & $g_2^R$ &  $g^R$\\
\hline
%\cline{1-6}
6 & 1.573(2) & 2.749(4) & 24(1) & 2.2(2) & -5.1(2) \\
\hline
8 & 1.572(2) & 2.785(2) & 27(2) & 2.6(4) & -4.8(2) \\
\hline
10 & 1.570(2) & 2.784(2) & 32(4) & 2.6(7) & -4.7(5) \\
\hline
\end{tabular}
%\end{center}
\caption{Correlation lengths and renormalized couplings at $\kappa_1=0.255$,
$\kappa_2=0.234$.}
\label{TABLE}
\end{table}

We checked the correctness of our measures by carrying out several tests in
some well-known models. In the first place we measured the renormalized
coupling in the one-scalar $\phi^4$ theory, using the same lattices, and
getting results fully compatible with those of ref.\cite{luscher}. In the
second place we simulated the coupled system, but taking
$g_1^{(0)}=g_2^{(0)}$ and $g^{(0)}=g_1^{(0)}/3$, which corresponds to the
O(2) theory, checking that these relations were also verified by the
renormalized parameters.

The results for our model are shown in table \ref{TABLE}. The correlation
lengths are stable with the lattice size, which indicates that we are using
big enough lattices. On the other hand, they are sufficiently large to
believe that we are in the
scaling region. The renormalized couplings are rather stable as well.
Though the large errors do not allow an accurate study of the thermodynamic
limit, it is clear that eq.(\ref{inst}) is verified by $g_2^R$ and $g^R$,
which is the essential thing. These renormalized quantities are all smaller
(in modulus in the case of $g^R$) than the bare ones
($g_1^{(0)} \simeq 125$, $g_2^{(0)} \simeq 4.9$, $g^{(0)} \simeq -7.5$),
as expected from
triviality, but while the values of $g_2^R$ and $g^R$ are rather small and
well inside the perturbative region for ISB, $g_1^R$ is quite large. In
order to get a significantly smaller value for  $g_1^R$ we should have
performed the measures much deeper in the scaling region, but this would
have been very difficult, because we would have needed much larger lattices
and because, closer to the double critical point C, the nearby strong first
order transition makes difficult to get clean results.

The results of the numerical measurements were then extrapolated deeper in
the scaling region using RG-equations, and taking the values of table
\ref{TABLE} as initial data for the numerical integration. The relevant
beta-functions were computed to two-loops order, keeping the corrections to
scaling for finite values of the correlation lengths up to one-loop. In view
of the large value of $g_1^R$ in the initial point, in the beta-function
for $g_1^R$, we included also the three loops contribution, of order
$(g_1^R)^4$. According to \cite{luscher}, this approximation should be
fully reliable for values of $g_1^R$ even larger than ours. Upon
integrating the RG equations along the line parallel to the $\kappa_1$ axis
passing through the simulation point, we found that while $g_1^R$ decreases
rapidly, eventually entering in the perturbative region, the evolution of
$g_2^R$ and $g^R$ is such that the perturbative condition for ISB
eq.(\ref{inst}) remains strongly satisfied. We thus concluded that in the
scaling region of our $(\kappa_1,\kappa_2)$ plane there were points for
which perturbation theory would have definitely predicted ISB for $\phi_2$.

Having analyzed the $T=0$ theory, we turned to the $T>0$ case. The issue
was to study the direction of the shift of the critical line $A$ as a
function of the temperature. For that we needed an accurate determination
of some critical points both on symmetric and on asymmetric lattices. Since
they had to be found with a high numerical precision, in order to clearly
distinguish the phase transitions corresponding to different values of
$N_t$, we could not afford to explore the entire $(\kappa_1,\kappa_2)$
plane, and thus we fixed once and for all the value of $\kappa_1=0.21$, and
searched on that vertical line for the accurate critical values of
$\kappa_2$, $\kappa_2^c(N_t)$, corresponding to the transition of $\phi_2$.
We will see that $\kappa_2^c(N_t)$ grows when $N_t$ diminishes, i.e. that the
disordered phase gets larger when increasing the temperature, indicating
the non existence of ISB.

Near the transition points, the correlation length for the field $\phi_1$
was about 0.7, not very large but still reasonable for us to believe that
we were probing the scaling region. We  simulated lattices with
$N_t=2,3,4,5,N_s$ and $N_s=6,8,10,12,14,16,20,24,28$. More exactly, for
$N_t=2,3,4,5,N_s$, we simulated up to ${N_s}^{max}=16,20,24,28,20$
respectively, making sure in all cases that the results showed asymptotic
behavior.
\begin{figure}[h!]
\epsfig{figure= 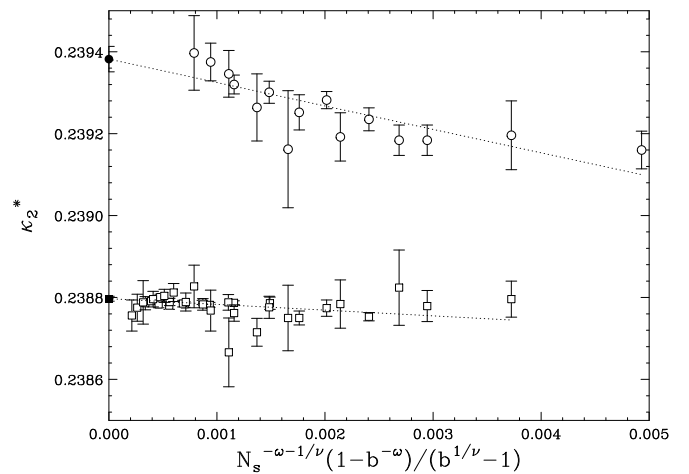,angle=90,width=250pt}
\caption{Fits to obtain $\kappa_2^{c}$ for $N_t=2,5$.}
\label{BINDER}
\end{figure}
For each of the points and lattices that were simulated
we made between 5 and 15 million iterations of the Metropolis algorithm.
The autocorrelation time $\tau_2$ ranged from 200 to 2,500 iterations.
In some runs we used also the cluster algorithm, but found that it was not
efficient for our values of the parameters.
The errors in the estimation of the observables were calculated
with the jackknife method.

The estimators used and the method followed to take the thermodynamic limit
were exactly the same as those used in ref. \cite{bitu} and we refer to it
the reader for the details. Having
measured the Binder cumulants, we extrapolated them in a narrow $\kappa_2$
interval around the simulation point using the spectral density method. The
values of $\kappa_2^c$ in the thermodynamic limit were obtained by looking
at the intersections $\kappa^*_2(N_{s1}, N_{s2})$ among all possible pairs
of curves $U_i(N_s,\kappa_2)$ and using the following scaling law
\be
\kappa^*_2(N_{s}, b N_{s})- \kappa_2^c =
\frac{1-b^{-\omega}}{b^{1/\nu}-1} N_s^{-\omega -1/\nu}~,\label{binsca}
\ee
where $\omega$ is the exponent for the corrections to scaling. In
fig.\ref{BINDER} we show the fits to the above expression for the $N_t=2,5$
lattices. For the calculation of the error, due account was taken of the
fact that the pair crossings were not all independent of each other.

For the asymmetric lattices, since the scaling parameter is $N_s$, while
$N_t$ is fixed, according to the hypothesis of dimensional reduction and
universality, we  used the exponents of the Ising model in {\it three}
dimensions, namely $\nu =0.63$ and $\omega=0.8$. As a check, we also
computed $\nu$ directly from the data obtaining values fully compatible
with the above one.

For the symmetric lattices we should use the mean field exponents
$\omega=0,\nu=1/2$. In this case, apart from logarithmic corrections, all
the crossings should occur for the same value of $\kappa_2$, and this is
approximately what happens. In order to give an estimate of $\kappa_2^c$,
we took the limit of the result for $\omega \to 0$, obtaining a very fast
convergence.

\begin{figure}[h!]
\epsfig{figure= 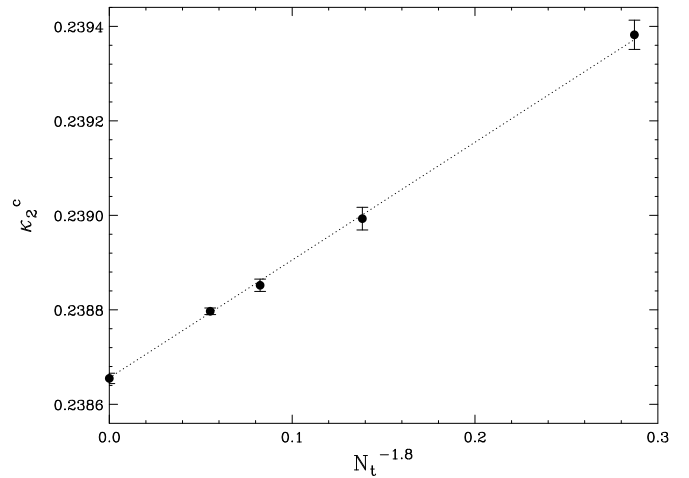,angle=90,width=250pt}
\caption{$\kappa_2^{c}$ as a function of $N_t^{-1.8}$.}
\label{K2_NT}
\end{figure}

In this way we have obtained, for each value of $N_t$, an estimation of
$\kappa_2^{c}(N_t)$ with $N_s \to \infty$.
This constitutes the most important result of the simulations and
is shown in fig.\ref{K2_NT}:
for a constant $\kappa_1$, starting from the symmetric lattices (the point
with $1/N_t=0$), $\kappa_2^{c}$ for the transition of the $\phi_2$ field
increases monotonically when $N_t$ is decreased; this means that the
critical points for $N_t$ finite shift deeper and deeper in the ordered
region of the $T=0$ model, i.e. raising the temperature disorders the
system, as it happens in normal cases (see for a comparison \cite{bitu})
and contrary to what is required for ISB to occur. In order to make sure
that the values of $N_t$ that we simulated were large enough, we checked
the scaling of $\kappa_2^{c}$ as a function of $N_t$. From FSS analysis
\cite{car}, one would in fact expect a scaling of the form
 $\kappa_2^{c}(N_t)- \kappa_2^{c}(\infty)\sim N_t^{-1/\nu}$, with
$\nu=1/2$. Letting $\nu$ variable, the best fit to the previous expression
is obtained for $1/\nu=1.8(2)$; this is the fit shown in the figure. This
makes us confident about the fact that no changes in the behavior of
$\kappa_2^{c}(N_t)$ should be seen for larger $N_t$'s.

In principle, our measurements are not conclusive, in that we cannot
exclude a priori the possibility that the $\phi_2$ transition lines of the
finite $N_t$ lattices move below that of the symmetric lattice for
values of $\kappa_1$ closer to $C$.
% In order to imply ISB this should occur
%though on the left of the double critical point C.
Unfortunately we could not obtain accurate results in a reasonable computer
time in that region, due to the existence of strong metastabilities
associated with the first order transitions. However, a toy-test on the
$4^4$ and $4^3 {\times} 2$ lattices near the point C confirmed the absence of
ISB.

In conclusion, we have found no trace of
ISB in our two-scalar
model, in a region of parameters for which
perturbation theory predicts this phenomenon.

%\section*{Acknowledgments}
\bigskip

We are grateful to the RTN collaboration and J.M. Carmona for discussions.
Work partially
supported by CICYT (AEN96-1670). D.I. and C.L.U.
acknowledge MEC and DGA for their fellowships.

\bigskip
\small $^*$ Electronic address: bimonte@napoli.infn.it

\small $^{\dagger}$ Electronic address: david, tarancon, clu@sol.unizar.es
%\section{References}

%\end{enumerate}

\end{document}